\documentclass[aps,prd,twocolumn,tightenlines,letterpaper,superscriptaddress,nofootinbib,floatfix]{revtex4-2}
\usepackage{amssymb,latexsym}
\usepackage{amsmath,amsbsy,bbm}
\usepackage{epsfig,bm,color}
\usepackage{graphicx}
\usepackage[normalem]{ulem}
\usepackage[dvipsnames]{xcolor}

\usepackage{calligra}

\DeclareMathAlphabet{\mathcalligra}{T1}{calligra}{m}{n}
\DeclareFontShape{T1}{calligra}{m}{n}{<->s*[2.2]callig15}{}

\newcommand{\bcr}{\bm{r}}

\begin{document}
%%%%%%%%%%%%%%%%%%%%%%%%%%%%%%%%

\def\a{{\alpha}}
\def\be{{\beta}}
\def\d{{\delta}}
\def\D{{\Delta}}
\def\P{{\Pi}}
\def\p{{\pi}}
\def\e{{\varepsilon}}
\def\ep{{\epsilon}}
\def\l{{\lambda}}
\def\L{{\Lambda}}
\def\m{{\mu}}
\def\o{{\omega}}
\def\O{{\Omega}}
\def\S{{\Sigma}}
\def\t{{\tau}}
\def\x{{\xi}}
\def\X{{\Xi}}
\def\z{{\zeta}}

\def\ol#1{{\overline{#1}}}
\def\c#1{{\mathcal{#1}}}
\def\b#1{{\bm{#1}}}
\def\eqref#1{{(\ref{#1})}}

%%%%%%%%%%%%%%%%%%%%%%%%%%%%%%%%%%%
%%%%%%%%%%%%%%%%%%%%%%%%%%%%%%%%%%%
\author{Johannes Kirscher}
\affiliation{
Theoretical Physics Division, School of Physics and Astronomy, The University of Manchester, Manchester M13 9PL, UK}

\author{Brian~C.~Tiburzi}
%\email[]{$\texttt{btiburzi@ccny.cuny.edu}$}
\affiliation{
Department of Physics,
        The City College of New York,
        New York, NY 10031, USA}
\affiliation{
Graduate School and University Center,
        The City University of New York,
        New York, NY 10016, USA}

%%%%%%%%%%%%%%%%%%%%%%%%%%%%%%%%%%%%
%%%%%%%%%%%%%%%%%%%%%%%%%%%%%%%%%%%% 
\title{
Two Particles with Zero-Range Interaction in a Magnetic Field
} 
%%%%%%%%%%%%%%%%%%%%%%%%%%%%%%%%%%%%
%%%%%%%%%%%%%%%%%%%%%%%%%%%%%%%%%%%%

\begin{abstract}
Energy levels are 
investigated
for two charged particles 
possessing
an attractive, 
momentum-independent, 
zero-range interaction in a uniform magnetic field. 
A transcendental equation governs the spectrum, 
which is characterized by a collective Landau-level
quantum number incorporating both center-of-mass and relative degrees of freedom. 
Results are obtained for a system of one charged and one neutral particle, 
with the interaction chosen to produce a bound state in vanishing magnetic field.
Beyond deriving the weak-field expansion of the energy levels, 
we focus on non-perturbative aspects. 
In the strong-field limit, 
or equivalently for a system in the unitary limit,
a single bound level with universal binding energy exists. 
By contrast, 
excited states are resonances that disappear into the continuum as the magnetic field is raised beyond critical values.
A hyperbola is derived that approximates the number of bound levels as a function of the field strength remarkably well.
\end{abstract}

\maketitle

%%%%%%%%%%%%%%%%%%%%%%%%%%%
\section{Introductory Remarks}
%%%%%%%%%%%%%%%%%%%%%%%%%%%

The study of quantum systems in external fields is central to many disciplines,
with applications ranging from the study of ultracold gases of trapped atoms%
~\cite{Bloch:2008zzb}
to quantum field theories in condensed matter, nuclear, and particle physics%
~\cite{Miransky:2015ava}.
In numerical gauge-theory calculations, 
external fields present a practical 
computational
method to study properties 
of light nuclei using lattice quantum chromodynamics (LQCD).
For an overview, 
see%
~\cite{Detmold:2004qn,Beane:2010em,Davoudi:2020ngi}
and references therein. 
The first LQCD results for light nuclei in external magnetic fields%
~\cite{Detmold:2015daa,Beane:2015yha,Chang:2015qxa}
point to an intriguing possibility. 
The energies of a subset of two-nucleon systems in large magnetic fields appear to approach binding thresholds, 
which suggests an analogue of a Feshbach resonance%
~\cite{Feshbach:1958nx,Fano:1961zz,Feshbach:1962ut}. 
While such resonances are extensively studied in atomic systems%
~\cite{RevModPhys.82.1225}, 
the LQCD results hint at a similar effect in two-nucleon systems. 
The magnetic field, 
moreover, 
could provide a lever with which to tune two-nucleon systems simultaneously to threshold, 
arriving at a point of conformal symmetry%
~\cite{Braaten:2003eu}. 

Unlike atomic systems, 
the magnetic fields required to alter strongly interacting systems are exceptionally large, 
$\c O(10^{19})$ 
Gauss. 
Fields of this size
are routine in LQCD computations,  
due to quantization conditions required of uniform fields on a torus%
~\cite{AlHashimi:2008hr}. 
In astrophysics, 
strong magnetic fields have been conjectured to occur in the interiors of magnetars%
~\cite{Duncan:1992hi,Broderick:2000pe,Harding:2006qn}.
Such extreme fields, 
have been estimated to exist in non-central heavy-ion collisions%
~\cite{Skokov:2009qp,McLerran:2013hla}, 
moreover,
for which considerable work has explored the implications for QCD matter%
~\cite{Kharzeev:2013jha}.

In the present work, 
we investigate a two-body problem in uniform magnetic fields. 
For atomic systems, 
there is substantial work on Coulomb interactions in an external magnetic field, 
\emph{e.g.}~\cite{Ruder:1994}.
The analogous study of QCD bound states has begun recently; 
examples include
quarkonium in external magnetic fields%
~\cite{Alford:2013jva,Bonati:2015dka}, 
as well as modification of the quark-antiquark potential computed from LQCD%
~\cite{Bonati:2014ksa}. 
For two protons in a magnetic field, 
the center-of-mass and relative motion can be separated.  
In this system, 
the effect of a sufficiently large magnetic field has been proven to result in a bound di-proton state%
~\cite{Allor:2006zy}.
The argument hinges on magnetic confinement of the protons into narrow cyclotron orbits. 
In such a scenario, 
the short-range nuclear interaction results in an attractive one-dimensional potential well, 
which always supports a bound state.%
\footnote{In atomic systems, 
a related argument was made to prove the existence of bound negative ions in large magnetic fields%
~\cite{PhysRevLett.39.1068}.
For this case, 
the atomic polarizability leads to the attractive interaction with the electron.} 
To address the neutron-proton system in magnetic fields, 
however, 
one needs to confront the non-separability of the center-of-mass and relative degrees of freedom, 
and we take a first step in this work.

General results for charged particles with translationally invariant interactions in a magnetic field were established in%  
~\cite{AVRON1978431}. 
In particular, 
a quantum number exists that collectively incorporates the center-of-mass and relative degrees of freedom, 
and each level is infinitely degenerate. 
This provides a striking generalization of the Landau problem to all values of the magnetic field. 
We exhibit these results by solving for the energy levels of two charged particles interacting with a contact potential. 
The choice of an attractive, 
momentum-independent, 
zero-range interaction is motivated by the pion-less effective
%field 
theory of few-nucleon systems%
~\cite{Beane:2000fx,Bedaque:2002mn,Epelbaum:2008ga,Hammer:2019poc}.%
\footnote{
Throughout this initial study, 
we assume the particles are scalars. 
Treatment of the two-nucleon problem requires 
including spin and magnetic moments, 
which is left to future work. 
}
The zero-range interaction between particles produces valuable simplifications that facilitate determining the spectrum.  
The energies of this system are found to obey a transcendental equation;
moreover, 
through a judicious form of renormalization, 
the inverse solution  
(magnetic field as a function of energy) 
is obtained simply by quadrature.

We begin our treatment with a general discussion in 
Sec.~\ref{sec:GenDis}, 
where we exhibit the quantum number incorporating both center-of-mass and relative degrees of freedom. 
In Sec.~\ref{sec:Example}, 
we focus on a system consisting of one charged and one neutral particle, 
for which the spectrum is obtained as a function of the magnetic field. 
Perturbative expansions are derived about the small-field and unitary limits.
The number of bound levels supported as a function of the magnetic field is shown to obey a simple equation. 
A summary of results is given in 
Sec.~\ref{summy}, 
while various technical details are provided in 
Apps.~\ref{sec:AppA} 
and 
\ref{sec:AppB}.

%%%%%%%%%%%%%%%%%%%%%%%%%%%
\section{General Discussion}                                     
%%%%%%%%%%%%%%%%%%%%%%%%%%%
\label{sec:GenDis}

The system under consideration is that of two particles 
with charges 
$q_a$
and masses 
$m_a$, 
where 
$a= 1$, $2$. 
The particles are in a uniform magnetic field 
and 
interact through an attractive, zero-range potential in three spatial dimensions. 
The Hamiltonian for this system has the form
\begin{equation}
H
=
H^{(0)} - c \, \d(\bcr)
\label{eq:H}
,\end{equation} 
where
$\bcr = \b r_1 - \b r_2$
is the relative coordinate between the particles.
Without the two-body potential,%
\footnote{ 
A regulator is needed to define the zero-range
potential in Eq.~\eqref{eq:H};
hence, 
$c$ is a running coupling. 
We demonstrate in Eqs.~\eqref{eq:cL1} and \eqref{eq:cL} that
the running of 
$c$ 
determined in zero magnetic field
yields regulator-independent equations for the spectrum even when
the field is present; {\it i.e.}, the magnetic field does not affect the interaction
at zero range.
}
the Hamiltonian for two particles in a magnetic field is additive
\begin{equation}
H^{(0)}
=
\sum_{a=1}^2
H^{(0)}_a
=
\sum_{a=1}^2
\frac{\big[ \b p_a - q_a \b A(\b r_a) \big]^2}{2 m_a} 
\label{eq:H0}
,\end{equation}
where 
$\b A$
is the vector potential that specifies the magnetic field. 
For arbitrary values of the charges and masses of the two particles, 
there is no general separation between the relative and center-of-mass motion.%
\footnote{
Solutions for the separable case of equal charge-to-mass ratios do
emerge as a limit of the general treatment above, 
see Footnote 9. 
A simpler method for solving such systems, 
however, 
is to enforce factorization of the relative and center-of-mass wave-functions from the outset. 
Additionally, 
the special case where the total charge of the system vanishes requires a separate treatment.
}
At the quantum level, 
the eigenvalues of 
$H$
are neither the sum of relative and center-of-mass contributions,
nor are they the sum of single-particle contributions. 

A bound state with energy 
$\c E$
appears as a solution to the homogeneous Lippmann-Schwinger integral equation
\begin{eqnarray}
\Psi%_N 
(\b r_1, \b r_2 )
= 
- c
\int d \b R \, G^{(0)} (\b r_1, \b r_2; \b R, \b R | \c E)
\Psi%_N 
(\b R, \b R )
,\,\,
\label{eq:Schr}
\end{eqnarray}
where 
$G^{(0)}$
is the non-interacting two-particle Green's function 
\begin{eqnarray}
G^{(0)} (\b r'_1, \b r'_2; \b r_1, \b r_ 2 | \c E)
=
\big\langle \b r'_1, \b r'_2 \big| \frac{1}{\c E - H^{(0)} + i \ep} \big| \b r_1, \b r_ 2 \big\rangle
\label{eq:G0}
,\quad\,\,\,\end{eqnarray} 
and the integration is over the center-of-mass coordinate 
$\b R$.
By virtue of the contact potential, the integration over the relative separation
amounts to an evaluation of the wave-function in the integrand at 
$\bcr=0$. 
While center-of-mass and relative coordinates are not completely decoupled in 
$H^{(0)}$,  
partial factorization of the wave-function is possible. 
Assuming a non-vanishing charge for the system
$Q \equiv q_1 + q_2$,
the center-of-mass wave-function is specified by the momentum parallel to the magnetic field, along with one other quantum number.%
\footnote{
More formally, 
we can simultaneously specify two components of the 
so-called 
pseudo-momentum~%
\cite{AVRON1978431}. 
Parallel to the magnetic field, 
the pseudo-momentum is simply the total momentum
$P_z$.  
The $x$-component of pseudo-momentum is 
$- \sqrt{QB} \, \xi$, 
where
$\xi$
appears in 
Eq.~\eqref{eq:xis}.
}
This quantum number can be chosen as an additional component of the total momentum,
for which the asymmetric gauge is natural. 
Therefore, 
we adopt
\begin{equation}
\b A (\b r)= \left( - B y, 0, 0 \right)
\label{eq:A}
,\end{equation} 
leading to the magnetic field
$\b B = B \bm{\hat{z}}$.
Accordingly, 
the $x$- and $z$-components of the total momentum 
$\b P= \b p_1 + \b p_2$,
are good quantum numbers. 
For any vector
$\b v$, 
we define a related vector
$\bm{\widetilde{v}}$,
with two non-vanishing components 
$\bm{\widetilde{v}}
= \left( v_x, 0, v_z \right)
$,
so that, 
\emph{e.g.},  
$\bm{\widetilde{P}}$
refers to the good components of $\b P$.

The two-particle wave-function can be written in the partially factorized form
\begin{equation}
\Psi (\b r_1, \b r_2)
=
e^{i \bm{\widetilde{P}} \cdot \bm{\widetilde{R}}} \,\,
\Psi_{P_x} (Y, \bcr)
,\end{equation}
in which 
$\Psi_{P_x} (Y, \bcr)$
reflects the coupled center-of-mass and relative motion.  
The 
$\bm{\widetilde{P}}$-projected 
non-interacting Green's function is obtained through the Fourier decomposition
\begin{eqnarray}
G^{(0)} (\b r'_1, \b r'_2; \b r_1, \b r_2 | \c E) 
&=&
\int \frac{d\bm{\widetilde{P}}}{(2\p)^2} \,
e^{i \bm{\widetilde{P}} \cdot \left( \bm{\widetilde{R}}' - \bm{\widetilde{R}}\right)}
\qquad
\qquad
\notag \\
&& \phantom{sp} \times
G^{(0)}_{P_x}(Y', \bcr';  Y, \bcr | E )
,\,\,\label{eq:G0proj}
\end{eqnarray}
where the shifted energy 
$E = \c E - \frac{P_z^2}{2M}$
serves to remove all 
$P_z$-dependence
from the projected Green's function.
Projecting 
Eq.~\eqref{eq:Schr}
onto states of good 
$\bm{\widetilde{P}}$, 
leads to the wave equation
\begin{eqnarray}
\Psi_{P_x} (Y, \bcr)
= 
- c
\int dY' G^{(0)}_{P_x} (Y, \bcr; Y', \b 0 | E)
\Psi_{P_x} (Y', \b 0 )
.\quad\,\,\,
\label{eq:SchrYr}
\end{eqnarray}
As with 
Eq.~\eqref{eq:Schr}, 
the wave-function appearing in the integrand is evaluated at zero relative separation owing to the contact interaction.
Knowledge of this restricted wave-function
$\Psi_{P_x} (Y', \b 0 )$
thus determines the value for 
$\bcr \neq \b 0$
by convolution with the non-interacting Green's function.

The restricted equation 
\begin{eqnarray}
\Psi_{P_x} (Y, \b 0)
= 
- c
\int dY' G^{(0)}_{P_x} (Y, \b 0; Y', \b 0 | E)
\Psi_{P_x} (Y', \b 0 )
,\quad\quad
\label{eq:SchrYr0}
\end{eqnarray}
is obtained by evaluating 
Eq.~\eqref{eq:SchrYr}
at
$\bcr = \b 0$, 
and suffices to determine the energy 
$E$. 
Simplifications arise because the configuration space is restricted to both particles being at the same point, 
for which there remains only center-of-mass motion of a system having charge 
$Q$. 
This is manifest in the functional form of the restricted non-interacting Green's function 
\begin{equation}
G^{(0)}_{P_x} (Y, \b 0; Y', \b 0 | E)
=
\sqrt{QB} \, \,
g^{(0)} (\x; \x' | E)
\label{eq:Gred}
,\end{equation}
which has a simple form in terms of the dimensionless center-of-mass coordinates
\begin{equation}
\x = \sqrt{QB} \, (Y-Y_0)
\quad \text{and} \quad
\x' = \sqrt{QB} \, (Y'-Y_0)
\label{eq:xis}
,\end{equation}
where 
$Y_0 = \frac{P_x}{QB}$
is the 
$y$-coordinate 
of the center of cyclotron motion of the restricted system. 
Written as a function of the variables
$\x$ and $\x'$, 
the restricted Green's function 
$g^{(0)}$
has no dependence on 
$P_x$. 
Consequently, 
the energy determined from 
Eq.~\eqref{eq:SchrYr0} 
is independent of 
$P_x$,
which confirms that each 
$E$ 
is infinitely degenerate.

For arbitrary charges and masses of the two particles, 
explicit computation establishes that 
$g^{(0)}$ 
is diagonal in the basis of Hermite functions%
\footnote{
The dimensionless Hermite functions employed above form an orthonormal basis, 
where the $n$-th member is given by 
$$
\psi_n(\x)
=
H_n(\x) \,
e^{-\frac{1}{2} \x^2}
\Big/\sqrt{2^n \, n! \sqrt{\p}}
,$$
with
$H_n(\x)$
as the corresponding Hermite polynomial. 
}
\begin{equation}
g^{(0)} (\x; \x' |  E)
=
\sum_{n=0}^\infty
g^{(0)}_{n} (E) \,
\psi_n (\x)
\, 
\psi_n (\x')
\label{eq:LL}
.\end{equation}
Projecting
Eq.~\eqref{eq:SchrYr0}
onto the $n$-th Hermite function, 
we obtain the equation 
\begin{equation}
\frac{1}{c} + g^{(0)}_{n} (E) = 0
\label{eq:eigval}
,\end{equation}
that determines 
$E$
for each value of 
$n$. 
This quantum number is the Landau level of the center of mass only when the two particles are restricted to the same point. 
With 
$\bcr \neq \b 0$, 
the Green's function 
$G^{(0)}_{P_x} (Y, \bcr; Y', \b 0 | E)$
is not diagonal in the basis of Hermite functions; 
accordingly, 
factorization of 
$\Psi_{P_x} (Y, \bcr)$
does not result from 
Eq.~\eqref{eq:SchrYr}.
This reflects that the quantum number 
$n$
incorporates collective behavior of both center-of-mass and relative degrees of freedom.

%%%%%%%%%%%%%%%%%%%%%%%%%%%
\section{Determining the Energies: \\ An Example}      
%%%%%%%%%%%%%%%%%%%%%%%%%%%
\label{sec:Example}

As an example, 
we solve for the spectrum of a system consisting of one charged and one neutral particle.
Accordingly, 
we take 
$q_1 = Q$
and
$q_2 = 0$.
To exhibit various properties of the solution, 
we keep the masses
$m_1$
and
$m_2$
different.

%%%%%%%%%%%%%%%%%%%%%%%%%%%%%%%%%%%%%%%
\subsection{Equations for the Spectrum}
%%%%%%%%%%%%%%%%%%%%%%%%%%%%%%%%%%%%%%%

For this system, 
computation of the restricted, 
non-interacting Green's function, 
Eq.~\eqref{eq:Gred},
is detailed in 
App.~\ref{sec:AppA}. 
The calculation produces the coefficients 
$g_n^{(0)}(E)$
of
Eq.~\eqref{eq:LL}
expressed in terms of a Laplace transform. 
The equation for the bound-state energy, 
Eq.~\eqref{eq:eigval},  
thus takes the form
\begin{eqnarray}
\frac{1}{c(t_0)}
-
\int_{t_0}^\infty dt \, e^{E t} \, g^{(0)}_n(t)
=
0
\label{eq:AnswerUn}
,\end{eqnarray}
where we introduce a small-time cutoff 
$t_0$
to regulate the divergence at 
$t=0$.%
\footnote{
Identical results are obtained using more conventional regularization schemes, 
such as dimensional regularization
(using $d-2$ spatial dimensions longitudinal to the magnetic field)
with the power-divergence subtraction scheme%
~\cite{Kaplan:1998tg,Kaplan:1998we}.
}
In Eq.~\eqref{eq:AnswerUn}, 
the 
coefficients are
\begin{eqnarray}
g^{(0)}_n(t)
&=&
\left(\frac{\m}{2 \pi t}\right)^{\frac{3}{2}}
\frac{M}{m_1}
\, \frac{\sinh \frac{\Omega(t) \, t}{2}}{\sinh \frac{\o t}{2}}
\, e^{ - \left(n+\frac{1}{2} \right) \O(t) \, t} 
\label{eq:tdepLL}
,\end{eqnarray}
and have been written in terms of the charged particle's cyclotron frequency 
$\o = Q B / m_1$, 
as well as a time-dependent frequency 
$\Omega(t)$
defined through the relation%
\begin{equation}
\coth \frac{\O(t) \, t}{2}
=
\coth \frac{\o t}{2} + \frac{m_2}{m_1} \left( \frac{\o t}{2} \right)^{-1}
\label{eq:O}
.\end{equation}
This time-dependent frequency
is a curious feature governing
the restricted cyclotron motion of the center of mass.

To renormalize Eq.~\eqref{eq:AnswerUn},
we inspect the small-time behavior of the integrand. 
In this limit, 
there is no magnetic field dependence, 
and we find the zero-field result
\begin{equation}
\frac{1}{c(t_0)}
-
\int_{t_0}^\infty dt \,  e^{ E_0 t} \left( \frac{\m}{2 \pi t} \right)^{\frac{3}{2}}
=
0 \label{eq:cL1}
,\end{equation}
where 
$E_0 <0$
denotes the energy in zero magnetic field.
The 
$t_0\to 0$ 
divergence of the integral in 
Eq.~\eqref{eq:AnswerUn}
thus subtracts the identical divergence in 
$\frac{1}{c(t_0)}$.
A renormalized form of Eq.~\eqref{eq:AnswerUn} defined for
$t_0 =0$
%all $t$
%without a short-time cutoff 
is therefore 
%given by
\begin{eqnarray}
\int_0^\infty dt
\left[
e^{ E t}
\, g^{(0)}_n(t)
-
e^{E_0 t}
\left( \frac{\m}{2 \pi t} \right)^{\frac{3}{2}}
\right]
=0
\label{eq:BSE}
.\end{eqnarray}
%having safely taken 
%$t_0 \to 0$. 
Any real energy 
$E$ 
below the two-particle dissociation
threshold, 
see Eq.~\eqref{eq:E}, 
that solves this equation
corresponds to an infinitely degenerate bound state with
quantum number 
$n$.
Below, we utilize this equation to obtain the small-field expansion.

An alternate equation for the spectrum is obtained by realizing that 
Eq.~\eqref{eq:cL1} 
can be rewritten as
\begin{eqnarray}
\frac{1}{c(t_0)}
-
\int_{t_0}^\infty dt \,  e^{ E t} \left( \frac{\m}{2 \p t} \right)^{\frac{3}{2}}
=
\frac{\m}
{2 \p}
\left[
\sqrt{- 2 \m E} 
- 
\sqrt{ - 2 \m E_0} 
\right]
\notag \\
\label{eq:cL}
\end{eqnarray}
with 
$E$
corresponding to a general energy, 
and small
$t_0$
understood for the integral. 
Using 
Eq.~\eqref{eq:cL}
for the running coupling, 
the renormalized and analytically continued equation for the energy 
$E$
can be cast in the form
\begin{eqnarray}
b^{-\frac{1}{2}}
=
\c I_n (\chi, x) 
\label{eq:BSE2}
,\quad
\end{eqnarray}
where the function
$\c I_n (\chi, x)$
is defined in App.~\ref{sec:AppB}. 
Appearing above, 
$x$
is the mass ratio 
$x = m_1 / M$, 
the magnetic field in dimensionless units is defined as
\begin{equation}
b = \frac{QB}{-2 \m E_0}
\label{eq:b}
,\end{equation} 
and the energy 
$E$ 
is written in terms of 
$\chi(b)$
in the form
\begin{equation}
E 
= 
\frac{\o}{2} - \chi \frac{QB}{2\mu}
\label{eq:E}
.\end{equation} 
This parametrization of the energy
accounts for the fact that the threshold for two-particle decay is set by the energy of the lowest Landau level of the charged particle. 
With the sign of the second term in 
Eq.~\eqref{eq:E}, 
$\chi$
is the binding energy in units of 
$\frac{QB}{2\mu}$. 
Notice that 
Eq.~\eqref{eq:BSE2} 
readily gives the inverse function 
$b(\chi)$ 
by simply evaluating 
$\c I_n (\chi,x)$, 
which is a one-dimensional integral. 
In accordance with regularization independence, 
the energies determined numerically from 
Eqs.~\eqref{eq:BSE} 
and 
\eqref{eq:BSE2} 
agree.

%%%%%%%%%%%%%%%%%%%%%%%%%%%
\subsection{Small-Field Limit}                                    
%%%%%%%%%%%%%%%%%%%%%%%%%%%

For small magnetic fields, 
one can treat 
$\o t \ll 1$,%
\footnote{
This treatment seems invalid when 
$t \gtrsim \omega^{-1}$,
however, 
the integrand appearing in 
Eq.~\eqref{eq:BSE} 
is exponentially small in this domain. 
}
for which the time-dependent frequency in 
Eq.~\eqref{eq:O}
becomes approximately time independent and equal to the cyclotron frequency 
$\O$ 
of the center of mass
\begin{equation}
\O(t) \approx \frac{QB}{M} \equiv \O  
\label{eq:Omega}
.\end{equation} 
Retaining only this linear magnetic field dependence, 
the equation for the energies, 
Eq.~\eqref{eq:BSE},
becomes
\begin{equation}
\int_0^\infty \frac{dt}{t^{3/2}}
\left(
e^{\left[ E - (n+\frac{1}{2}) \O \right] t}
-
e^{E_0 t}
\right) 
=
0
\label{eq:smallB}
,\end{equation}
and leads to the requirement 
$E = E_0 + (n+\frac{1}{2}) \O$. 
For small magnetic fields, 
the energies are ordinary Landau levels of the center of mass.

%%%%%%%%%%%%%%%%%%%%%%%%%%%%%%%%%%%%%
\begin{figure}
\begin{center}
%%%%%%%%%%%%%%%%%%%%%%%%%%%%%%%%%%%%%
\includegraphics[scale=0.925]{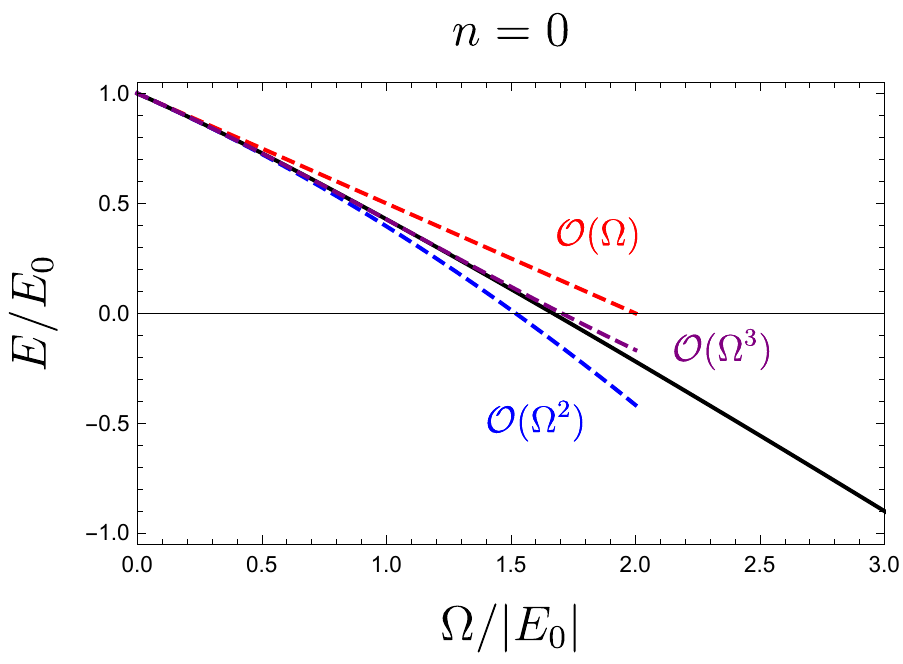}
%%%%%%%%%%%%%%%%%%%%%%%%%%%%%%%%%%%%%
\caption{Behavior of the lowest level in small magnetic fields.
For the case of equal masses, 
the ratio of the 
$n{=}0$ 
energy level 
$E$
to that in vanishing magnetic field
$E_0$
is plotted 
(solid curve) 
as a function of the magnetic field, 
using 
the center-of-mass cyclotron frequency
$\Omega$ 
in 
Eq.~\eqref{eq:Omega}.
Also plotted (dashed curves) are the perturbative approximations in powers of 
$\Omega / E_0$
from Eq.~\eqref{eq:smallBexp}. 
While the state becomes less bound, 
note that 
$E=0$
does not correspond to threshold, 
see Eq.~\eqref{eq:E}. 
}
\label{f:smallB}
%%%%%%%%%%%%%%%%%%%%%%%%%%%%%%%%%%%%%
\end{center}
\end{figure}
%%%%%%%%%%%%%%%%%%%%%%%%%%%%%%%%%%%%%

To obtain corrections to the leading-order result in small fields, 
we write 
$\D E = E - \left[ E_0 + (n + \frac{1}{2}) \O \right]$, 
and iteratively solve for 
$\D E$
using 
Eq.~\eqref{eq:BSE}. 
To third-order in the magnetic field, 
we find
\begin{eqnarray}
E
&=&
E_0
\Big[
1
+ 
\Big(n + \frac{1}{2} \Big) 
\frac{\O}{E_0} 
-
\frac{1}{48} \Big( 4 + \frac{m_2}{m_1} \Big) \frac{m_2}{m_1} 
\Big( \frac{\O}{E_0} \Big)^2
\notag \\
&&
\phantom{sp}-
\frac{1}{16}
\Big( n + \frac{1}{2} \Big) \frac{m_2}{m_1} 
\Big( \frac{\O}{E_0} \Big)^3
+
\c O(B^4)
\Big]
\label{eq:smallBexp}
.\end{eqnarray}
In particular, 
the term at second order is independent of the quantum number 
$n$,
and allows us to identify the magnetic polarizability of the bound state. 
Using the customary definition of the second-order energy shift
$\D E = - \frac{1}{2} 4 \p \be_M B^2 + \cdots$, 
we obtain the polarizability
\begin{equation}
\be_M  = \frac{\a}{24} \left( 4 + \frac{m_2}{m_1} \right)  \frac{m_2}{m_1} \, \frac{(Q/e)^2}{E_0 M^2} 
,\end{equation}
where 
$\a = e^2 / 4 \p$
is the fine-structure constant. 
The magnetic polarizability is diamagnetic 
$\be_M < 0$, 
due to the absence of paramagnetic spin contributions.

In Fig.~\ref{f:smallB}, 
we exhibit the small-field behavior of the 
$n = 0$
energy level
determined from Eq.~\eqref{eq:BSE2}. 
For simplicity, 
we take the equal mass case
$m_1 = m_2$. 
Additionally shown is the small-field behavior determined analytically from 
Eq.~\eqref{eq:smallBexp}. 
The perturbative expansion lines up well with the numerically determined energy, 
and appears to work beyond the 
$\Omega / |E_0| \ll 1$ 
regime.

%%%%%%%%%%%%%%%%%%%%%%%%%%%
\subsection{Unitary Limit}  
%%%%%%%%%%%%%%%%%%%%%%%%%%%

To determine the behavior of the binding energy in the unitary limit
$E_0 \to 0$, 
we turn to 
Eq.~\eqref{eq:BSE2}. 
With 
$b \gg 1$, 
the binding energy in the unitary limit
$\chi^{(0)}_\infty$
is obtained as the root of the equation
\begin{eqnarray}
\c I_n(\infty)
\equiv
\c I_{n} \Big(\chi^{(0)}_\infty, x \Big)= 0
\label{eq:Einfinity}
.\end{eqnarray}
For 
$n > 0$, 
there are no roots;
only the 
$n{=}0$
level is bound for a system at unitarity. 
The binding energy in this regime, 
moreover, 
is 
independent of the short-range structure of the interaction.
The bound state is thus a universal feature, albeit, in a
limited sense. 
In dimensionless units, 
the unitary-limit binding energy still depends upon the mass ratio between the charged and neutral particles. 
This mass-ratio dependence is a consequence of the coupling between relative and center-of-mass motion.
For identical masses ($x{=}\tfrac{1}{2}$), 
we obtain the numerical value
$\chi^{(0)}_\infty = 0.081588$.
Only when the charged particle is much more massive than the neutral one
($x{\to}1$)
does the system become unbound
$\chi^{(0)}_\infty = 0$.
In the opposite limit, 
the binding energy of a system with 
a relatively light charge 
($x{\to}0$)
attains, by contrast, its largest
value 
$\chi_\infty^{(0)} = 0.605444$, 
which is twice the root of the Hurwitz zeta-function, 
see Eq.~\eqref{eq:In}. 
Curiously, 
this value also appears as the binding energy of the spin-zero di-proton system in a strong magnetic field%
~\cite{Allor:2006zy}.%
\footnote{
We remark that the spectrum of that system near unitarity can be determined from 
$b^{-\frac{1}{2}} = - \frac{1}{\sqrt{2}} \zeta \left( \frac{1}{2}, \frac{\chi}{2} \right)$, 
which happens to be the 
$x \ll 1$
limit of 
Eq.~\eqref{eq:BSE2}.
}
In other words, a single charge that interacts resonantly with a fixed force center
(our model for $x{\to}0$) 
has the same ground-state energy as two 
particles with equal charge-to-mass ratios
that 
interact
via long-range Coulomb 
repulsion
in addition to the
zero-range attraction.

%%%%%%%%%%%%%%%%%%%%%%%%%%%%%%%%%%%%%
\begin{figure}
\begin{center}
%%%%%%%%%%%%%%%%%%%%%%%%%%%%%%%%%%%%%
\includegraphics[scale=0.925]{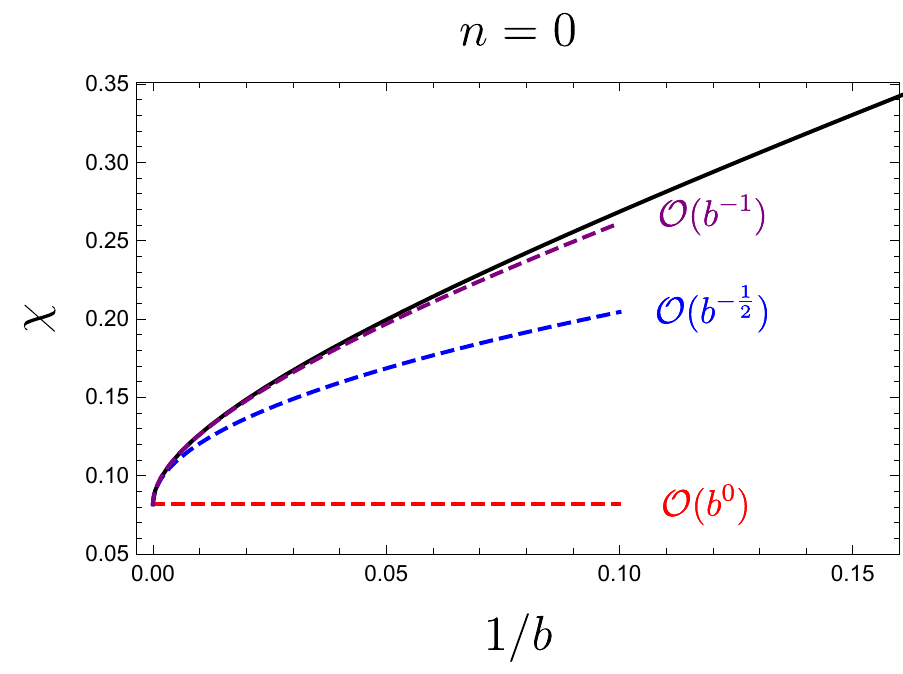}
%%%%%%%%%%%%%%%%%%%%%%%%%%%%%%%%%%%%%
\caption{Behavior of the lowest level near unitarity.
For the case of equal masses, 
$\chi$,
the binding energy in units of $\frac{QB}{2\m}$,
is plotted 
(solid curve) 
as a function of the inverse magnetic field, 
using 
$b$
defined in 
Eq.~\eqref{eq:b}.
Also plotted (dashed curves) are the successive approximations in
Eq.~\eqref{eq:largeBexp}
about the unitary limit 
$b \gg 1$. 
For this level, 
the threshold 
$\chi {=} 0$
is never reached. 
}
\label{f:largeB}
%%%%%%%%%%%%%%%%%%%%%%%%%%%%%%%%%%%%%
\end{center}
\end{figure}
%%%%%%%%%%%%%%%%%%%%%%%%%%%%%%%%%%%%%

Away from unitarity,
the binding energy is no longer universal as it depends on 
$E_0$
through the 
$b^{-\frac{1}{2}}$
corrections. 
To obtain these corrections, 
we write the binding energy in the expansion 
\begin{equation}
\chi 
= 
\chi^{(0)}_\infty 
+
\frac{1}{\sqrt{b}} \, \chi^{(1)}_\infty 
+ 
\frac{1}{b} \, \chi^{(2)}_\infty 
+ 
\c O \big(b^{-\frac{3}{2}} \big)
\label{eq:largeBexp}
,\end{equation}
and then solve for the higher-order terms using 
Eq.~\eqref{eq:BSE2} 
iteratively. 
This procedure results in
\begin{eqnarray}
\chi^{(1)}_\infty 
&=& 
1 \Big/ \c I'_0 (\infty)
=
0.388235
,\notag \\
\chi^{(2)}_\infty 
&=& 
- \tfrac{1}{2} \, \c I''_0 (\infty) \Big/ [ \c I'_0 (\infty)]^3
=
0.567853
,\quad\end{eqnarray}
where derivatives of 
$\c I_n(\chi,x)$
are with respect to 
$\chi$,
and the evaluation is at the location of the root in 
Eq.~\eqref{eq:Einfinity}, 
namely at
$\chi = \chi^{(0)}_\infty$. 
The formulas apply to any mass ratio, 
however, 
the numerical values quoted are exclusively for the equal-mass case
($x{=}\frac{1}{2}$). 
As the corrections away from unitarity are positive,  
the magnetic field always increases the ground-state binding energy
and thus does not allow crossing to a virtual state.

In Fig.~\ref{f:largeB}, 
we show the approach to unitarity of the 
$n = 0$
energy level
determined from 
Eq.~\eqref{eq:BSE2}. 
For simplicity, 
we again take the equal-mass case. 
Additionally shown is the behavior near the unitary limit obtained from
Eq.~\eqref{eq:largeBexp}, 
which does well in describing the magnetic-field dependence of the lowest energy level in this regime.

%%%%%%%%%%%%%%%%%%%%%%%%%%%%%%%
\subsection{Binding Thresholds and the Number of Levels}  
%%%%%%%%%%%%%%%%%%%%%%%%%%%%%%%
\label{sec:Thresh}

The binding threshold is reached when 
$\chi {=} 0$. 
As remarked above, 
the lowest level does not reach threshold; 
it remains bound in the unitary limit, 
which can alternately be viewed as the large-field limit with 
$E_0$
held fixed. 
The excited levels, 
however, 
reach threshold at critical values of the magnetic field.%
\footnote{
Note that for all values of the magnetic field, 
excited levels are unstable to radiative decay to the ground level. 
Such decays require electrodynamics, 
which is beyond the scope of this investigation. 
}
According to Eq.~\eqref{eq:BSE2}, 
the threshold for single-particle production occurs when the magnetic field has the value 
\begin{eqnarray}
b_c = \big[ \c I_n (0,x) \big]^{-2}
\label{eq:bceq}
,\end{eqnarray}
provided 
$\c I_n(0,x) > 0$. 
For the first excited level 
and with equal masses for the particles, 
we find
\begin{equation}
b_c = 3.87554, 
\,\,\, \text{for} \, \, 
n{=}1 
\, \, \text{and} \, \, x {=} \tfrac{1}{2}
\label{eq:bc}
.\end{equation}

For the higher-lying levels, 
the critical value of the magnetic field decreases monotonically.
Consequently,
we have the single-particle decay threshold at
$b_c \ll 1$,
for 
$n \gg 1$. 
Viewed as a function of the level number 
$n$, 
the critical field 
$b_c(n)$
can be obtained analytically in the large-$n$ limit. 
The analysis detailed in 
App.~\ref{sec:AppB}
produces the approximate formula
\begin{equation}
b_c = \frac{4}{2 n -1}  \left[ 1 + \c O (n^{-2}) \right]
\label{eq:bcc}
,\end{equation}
for the equal mass case. 
Surprisingly, 
the formula works at 
$3\%$
already for the 
$n{=}1$
level. 
Results improve with increasing 
$n$;
the approximate critical field for 
$n {=} 2$
differs by 
$0.6\%$
from the exact value.

%%%%%%%%%%%%%%%%%%%%%%%%%%%%%%%%%%%%%
\begin{figure}
\begin{center}
%%%%%%%%%%%%%%%%%%%%%%%%%%%%%%%%%%%%%
\includegraphics[scale=0.925]{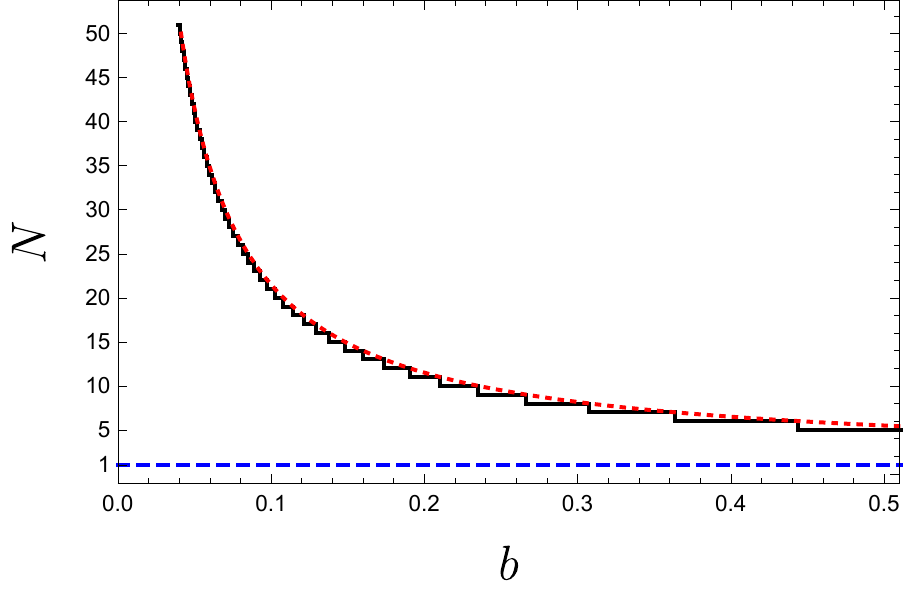}
%%%%%%%%%%%%%%%%%%%%%%%%%%%%%%%%%%%%%
\caption{
For the case of equal masses, 
the number of energy levels 
$N$ 
is plotted 
(solid curve) 
versus the magnetic field 
$b$ 
in 
Eq.~\eqref{eq:b}. 
Each step down is due to a level reaching the binding threshold. 
The ground level never reaches the single-particle decay threshold, 
for which a horizontal dashed line depicts the 
large-$b$ 
value
$N = 1$.
Additionally shown 
(dashed curve)
is the approximate function  
$N(b)$
given in 
Eq.~\eqref{eq:Nofb}. 
}
\label{f:number}
%%%%%%%%%%%%%%%%%%%%%%%%%%%%%%%%%%%%%
\end{center}
\end{figure}
%%%%%%%%%%%%%%%%%%%%%%%%%%%%%%%%%%%%%

The number of energy levels supported in a given magnetic field can be obtained from knowledge of the critical field for each value of 
$n$.
If we start at large
$b$, 
the number of levels is 
$N=1$
until the field is lowered below the value in 
Eq.~\eqref{eq:bc}. 
Below this value, 
the 
$n{=}1$
level has not crossed threshold, 
consequently
$N = 2$. 
Continuing to lower the magnetic field, 
successive critical fields will be crossed and the number of levels increased by one each time. 
The number of levels obtained by counting thresholds is shown in 
Fig.~\ref{f:number}
up to 
$N = 50$. 
By identifying 
$N = n + 1$ 
and simply inverting the formula for the critical field  
Eq.~\eqref{eq:bcc}, 
we obtain an approximate formula for the number of
levels as a function of the magnetic field
\begin{equation}
N = \tfrac{2}{b} + \tfrac{3}{2}    
\label{eq:Nofb}
.\end{equation}
The figure shows that this simple formula accounts
for the number of bound levels remarkably well.

%%%%%%%%%%%%%%%%%%%%%%%%%%%
\section{Summary}                                                     
%%%%%%%%%%%%%%%%%%%%%%%%%%%
\label{summy}

Systems of two interacting charged particles in a magnetic field present an analytically challenging non-separable problem. 
When the particles have an attractive, 
momentum-independent, 
zero-range 
interaction,
we confirm in 
Sec.~\ref{sec:GenDis}
that the bound states have infinite degeneracy, 
and are characterized by a quantum number 
that incorporates both center-of-mass and relative degrees of freedom. 
In small magnetic fields, 
this quantum number is the Landau level of the center of mass, 
as exhibited by 
Eq.~\eqref{eq:smallBexp}, 
which is the small-field expansion of the energy of a system consisting of one charged and one neutral particle. 
Near the unitary limit, 
there is dramatically different behavior for the ground and excited levels of this system. 
The lowest level is bound, 
with a universal binding energy determined by 
Eq.~\eqref{eq:Einfinity}. 
The excited states disappear into the continuum at critical values of the magnetic field, 
which are determined by 
Eq.~\eqref{eq:bceq}. 
The number of bound levels as a function of the magnetic field is well described by the remarkably simple formula in  
Eq.~\eqref{eq:Nofb}. 

A striking feature of this two-body system 
in a magnetic background is
that its bound-state spectrum is identical to
that of an effective one-body problem.
The effective interaction in this restricted space accounts for the
background field and zero-range interaction. 
The closed-form correlation function, 
Eq.~\eqref{eq:tdepLL}, 
is 
suggestive of an effective time-dependent Hamiltonian.
In this way,  
complexity of the non-separable system manifests
as non-trivial time-dependence 
%interaction 
of
the restricted system.
Nonetheless, renormalization can be carried out by hand, 
thus enabling straightforward determination
of spectral properties.
In particular, 
Eq.~\eqref{eq:BSE2} 
shows that the inverse of the solution, 
namely the magnetic field as a function of the binding energy, 
only requires evaluation of a one-dimensional integral. 
The number of systems possessing such almost-in-closed-form
solutions can be enlarged by generalizing and extending the problem solved here. 
Of particular interest would be a complete treatment of the two-nucleon system.

\begin{acknowledgments}
We thank M.~Birse for comments and discussions.
\end{acknowledgments}

%%%%%%%%%%%%%%%%%%%%%%%%%%%%%%%%%%%%%%%%%%%%
%%%%%%%%%%%%%%%%%%%%%%%%%%%%%%%%%%%%%%%%%%%%
%%%%%%%%%%%%%%%%%%%%%%%%%%%%%%%%%%%%%%%%%%%%
%%%%%%%%%%%%%%%%%%%%%%%%%%%%%%%%%%%%%%%%%%%%

\appendix

%%%%%%%%%%%%%%%%%%%%%%%%%%%
\section{Green's Function}
%%%%%%%%%%%%%%%%%%%%%%%%%%%
\label{sec:AppA}

The %time-independent 
Green's function
$G^{(0)}$
in Eq.~\eqref{eq:G0}
is computed from a Laplace transform 
using the inverse operator
\begin{equation}
\frac{1}{E - H^{(0)} + i \ep} 
= 
-
\int_0^\infty dt  \, e^{E t} \,  e^{ - H^{(0)} t}
\label{eq:Laplace}
,\end{equation}
where analytic continuation is necessary when
$E > 0$.
We refer to
$t$
as time; 
although, 
strictly speaking, 
it is imaginary time. 
As 
$H^{(0)}_1$
and
$H^{(0)}_2$
in Eq.~\eqref{eq:H0}
commute, 
$G^{(0)}$ 
factorizes into the product of single-particle Green's functions
\begin{eqnarray}
G^{(0)}
(\b r'_1, \b r'_2; \b r_1, \b r_2 | t)
&=&
\prod_{a=1}^2
\, \big\langle \b r'_a \big| e^{ -  H^{(0)}_a t} \big| \b r_a \big\rangle
\label{eq:Factorize}
.\end{eqnarray}
For the gauge chosen in Eq.~\eqref{eq:A},  
the Green's function of the charged particle takes the form
\begin{eqnarray}
\big\langle \b r'_1 \big| e^{ - H^{(0)}_1 t} \big| \b r_1 \big\rangle
&=&
\int \frac{d\bm{\tilde{p}}_{1}}{(2 \pi)^2} 
e^{ i \bm{\tilde{p}}_{1} \cdot (\bm{\tilde{r}}'_1 - \bm{\tilde{r}}_1 )}
e^{- \frac{p_{1,z}^2}{2 m_1} t}
\notag \\
&& \qquad \times 
\sqrt{QB} \, \,
\big\langle 
\x'_1, \, \o t \, \big| \x_1, 0 
\big\rangle
,\label{eq:Gsp}
\end{eqnarray}
and has been written in terms of the Green's function of the simple harmonic oscillator. 
With dimensionless conjugate variables
$\x$ and $p_\x$, 
the oscillator Hamiltonian appears as
$\c H = \frac{1}{2} \big( p_\x^2 + \x^2 \big)$,
for which the corresponding propagator we define as
$\langle \x', s \, | \x, 0 \rangle
=
\langle \x' | e^{- \c H \, s} | \x \rangle
$.
In Eq.~\eqref{eq:Gsp}, 
$\o$
is the charged particle's cyclotron frequency, 
and
the dimensionless charged-particle 
coordinates are %given by 
\begin{eqnarray}
\x'_1 = \sqrt{QB} \left( y'_1 - \frac{p_{1,x}}{Q B} \right)
\, \text{and} \,\,\,
\x_1 = \sqrt{QB} \left( y_1 - \frac{p_{1,x}}{Q B} \right)
.\notag \\
\end{eqnarray}
The coordinate-space Green's function of the neutral particle can be written as a Fourier transform
\begin{eqnarray}
\big\langle \b r'_2 \big| e^{ - H^{(0)}_2 t} \big| \b r_2 \big\rangle
&=&
\int \frac{d\b p_2}{(2\p)^3}
e^{i \b p_2 \cdot (\b r'_2 - \b r_2)} e^{ - \frac{\b p_2^2 }{2 m_2} t}
.\end{eqnarray}

Following Sec.~\ref{sec:GenDis}, 
we project 
Eq.~\eqref{eq:Factorize} onto good 
$\bm{\widetilde{P}}$,  
and restrict the computation of the matrix element to vanishing relative separation 
$\bcr' {=} \bcr {=} \b 0$.
The resulting time-dependent, 
non-interacting Green's function 
$g^{(0)} (\x; \x' | t)$
is both
$P_x$ independent
and diagonal in the basis of Hermite functions. 
The diagonal matrix elements are the 
$g_n^{(0)}(t)$
coefficients appearing in 
Eq.~\eqref{eq:tdepLL}.

%%%%%%%%%%%%%%%%%%%%%%%%%%%
\section{The Function $\c I_n(\chi,x)$}
%%%%%%%%%%%%%%%%%%%%%%%%%%%
\label{sec:AppB}

The function 
$\c I_n $ %(\chi, x)$
appearing in 
Eq.~\eqref{eq:BSE2}
is defined as
\begin{equation}
\c I_n(\chi,x)
=
-\tfrac{1}{\sqrt{2}} \,
\z \left( \tfrac{1}{2}, \tfrac{\chi+x}{2} \right) 
-
\c J_n(\chi,x)
\label{eq:In}
,\end{equation}
with 
$\z ( \frac{1}{2}, z)$
as a Hurwitz zeta-function, 
and where
\begin{eqnarray}
\c J_n(\chi, x)
&=&
\int_0^\infty ds \frac{e^{ (1-x- \chi) s}}{\sqrt{4 \pi} s^{3/2}}
\Big[
\,
\c F_n(x,s)
-
\frac{s}{\sinh s}
\Big]
.\quad\,
\label{eq:Jn}
\end{eqnarray}
Appearing in the above integrand is 
$\c F_n (x,s)$, 
which has the explicit form
\begin{equation}
\c F_n(x,s) 
= 
\frac{\sinh \frac{\Omega(s) \, s}{2}}{ x \, \sinh [(1-x)s]} 
\,
e^{- \left(n+ \frac{1}{2} \right) \Omega(s) \, s }
,\end{equation}
using 
$\coth \frac{\Omega(s) \, s}{2} = \coth [(1-x)s] + (xs)^{-1}$.
The integral defining
$\c J_n (\chi,x)$
converges in the infrared 
($s \to \infty$)
provided
$\chi \geq 0$. 
The definition of this function, 
moreover, 
is such that 
$\lim_{x \to 0} \c J_n (\chi, x) = 0$.

To investigate the critical field
$b_c$
in Eq.~\eqref{eq:bceq}
as a function of the level number, 
we differentiate 
Eq.~\eqref{eq:BSE2}
with respect to 
$n$
at 
$\chi =0$, 
which produces
\begin{equation}
\frac{1}{b_c^{3/2}} \frac{db_c}{dn}
=
\int_0^\infty ds \frac{e^{(1-x)s}}{\sqrt{\pi} s^{3/2}}
\frac{d}{dn} \c F_n(x,s)
.\end{equation}
Scaling 
$s \to 2 \, s / n$, 
and treating 
$n\gg 1$
leads to the approximate formula
\begin{equation}
\frac{1}{b_c^{3/2}} \frac{db_c}{dn}
=
-\int_0^\infty \frac{\, ds \, e^{- s}}{\sqrt{2 \pi n s}}
\, e^{\frac{s}{2n}}
\left[ 1 + \c O (n^{-2}) \right]
,\end{equation}
for the case of equal masses.
The exponential factor 
$e^{\frac{s}{2n}}$
need not be expanded in powers of 
$n^{-1}$, 
as the integral can be performed with this factor treated exactly. 
Solving the above differential equation for
$b_c$
subject to the condition 
$\lim_{n\to \infty} b_c (n) = 0$
results in 
Eq.~\eqref{eq:bcc}.

%%%%%%%%%%%
\bibliography{bibfile}
%%%%%%%%%%%

%%%%%%%%%%%
%%%%%%%%%%
%%%%%%%%%
\end{document}